\documentclass[showkeys,showpacs,preprint,amsmath,amssymb,epsf,aps]{revtex4}
\usepackage{dcolumn}
\usepackage{epsfig}

\usepackage{graphicx}% Include figure files
\usepackage{longtable}% Include figure files
\usepackage{dcolumn}% Align table columns on decimal point
\usepackage{bm}% bold math

\def\C60{C$_{60}$}

\def\afSOS {$\alpha_{SOS}$}
\def\afsos {$\alpha_{SOS}$\ }
\def\afFF {$\alpha_{FF}$}
\def\afFF {$\alpha_{FF}$}
\def\DHL {$\Delta_{HL}$}
\def\VF {\Bigg \vert_{\vec{F}=\vec{0}}}

\begin{document}

\title{Comparative study of unscreened and screened molecular static linear polarizability in 
the Hartree-Fock, hybrid-density functional, and density functional models}

\author{Rajendra R. Zope$^{1,2}$, Tunna Baruah$^{1}$, Mark R. Pederson$^3$,  and
B. I. Dunlap$^4$ }

\affiliation{$^1$Department of Physics, University of Texas at El Paso, El Paso, Texas 79959}

\affiliation{$^2$Department of Electrical and Computer Engineering, Howard university,
Washington, DC 20059}

\affiliation{$^3$Center for Computational Materials Science, Code 6392, Naval Research 
Laboratory, Washington DC 20375-5345}

\affiliation{$^4$Theoretical Chemistry Section, Naval Research Laboratory,
 Washington DC 20375-5345}

\date{\today}

%\pacs{36.40.Cg, 36.40.Mr, 36.40.Qv, 31.40.+z}
\pacs{31.15.Ar, 31.70.Hq, 34.70.+e, 84.60.Jt, 87.15.Mi}

%Use showkeys class option if keyword display desired
\keywords{polarizability, sum over states, finite field}

\begin{abstract}

        The {\em sum-over-states} 
(SOS)  polarizabilities are calculated within approximate mean-field electron theories 
such as the Hartree-Fock approximation and density functional models  using the eigenvalues 
and orbitals  obtained from the self-consistent solution of the single-particle equations.
The SOS polarizabilities are then compared with those calculated  using the finite-field (FF) method.
Three widely used mean-field models are used:
(1) the Hartree-Fock (HF) method, (2) the three parameter hybrid generalized gradient approximation
(GGA)
(B3LYP), and (3) the parameter-free generalized gradient approximation due to Perdew-Burke-Ernzerhof
(PBE).  The comparison is carried out for polarizabilities of 
142 molecules calculated using the 6-311++G(d,p) basis set
at the geometries optimized at the B3LYP/6-311G** level. 
The results show that the SOS method almost always overestimates the FF polarizabilities 
in the PBE and B3LYP models.  This trend is reversed in the HF method.  A few exceptions
to these trends are found.  The mean absolute errors (MAE) in the screened (FF) and unscreened 
(SOS) polarizability are  0.78 \AA$^3$,  
1.87 \AA$^3$,  and  3.44 \AA$^3$ for the HF, B3LYP and PBE-GGA methods, respectively. 
Finally, a simple scheme is devised to obtain FF quality polarizability from the SOS 
polarizability.

\end{abstract}

\maketitle

In this paper, we examine 
sum-over-states (SOS) and finite-field (FF) polarizabilities for Hartree-Fock (HF), a hybrid density-functional
B3LYP \cite{B3LYP_A,B3LYP_B}, and a generalized gradient approximation (GGA) pure density-functional,
Perdew-Burke-Ernzerhopf (PBE)\cite{PBE},
for a large set of molecules. These three are 
 popular 
independent-particle models that are routinely used to study the electronic structure of 
molecules.  We discuss the relationship between the SOS polarizabilities and the
one-electron eigenvalues of these mean-field methods.
 We then devise a simple empirical scheme to obtain 
finite-field quality estimate of the polarizability from the HF and PBE models.
We also investigate the differences in the predictions of polarizabilities 
from SOS and FF methods in the above popular single-particle methods.

    When a molecule is placed in a static electric field $\vec{F}$, the electronic charge 
of the  molecule redistributes.  This response is characterized by the polarizability 
of the molecule. 
 The molecular energy $E$ in the presence of a small  external field  can be expressed as a 
perturbative series in power of $\vec{F}$ as
\begin{eqnarray}
   E  & = &   E_0 + 
         \sum_{i}  \frac{\partial E}{\partial F_i}\VF F_i  + 
         \frac{1}{2!} \sum_{ij} \frac{\partial^2 E}{\partial F_i\partial F_j}\VF F_i F_j  \\
          \nonumber
        & & ~~+ 
          \frac{1}{3!} \sum_{ijk} \frac{\partial^3 E}{\partial F_i\partial F_j \partial F_k}\VF F_i F_j F_k 
           + ... \\
      & = &  E_0 +  \sum_i  \mu_i  F_i + \sum_{ij} \frac{1}{2} \alpha_{ij} F_i F_j + \sum_{ijk} \frac{1}{3!}
          \beta_{ijk} F_i F_j F_k + ..., 
          \label{eq:E_Taylor}
\end{eqnarray}
 where, $E_0$ is the total energy of the molecule in the absence of the external electric field, 
$\mu$ is the induced dipole moment, $\alpha$ is the 
linear polarizability, and  $\beta$ is the first hyper-polarizability; the indices i,j,k denote the 
Cartesean components. The calculation of the 
linear polarizability 
$\alpha_{ij}, $ 
which is the topic of the present manuscript, requires calculation of the second derivative of the 
total molecular energy with respect to applied field. 
The $\alpha_{ij}$ terms can be calculated using perturbation theory or 
by finite difference technique by using the total energies calculated at various field strengths.
A summary of various methods for practical calculations of polarizabilities can be found in 
several review articles\cite{Reviews,R2}. Using standard time-independent perturbation theory, 
one can obtain the well-known SOS expression 
(see, for example Bartlett and Sekino in Ref.\onlinecite{R2}).
The SOS expression for  a diagonal element of the polarizability tensor is given by 
\begin{equation}
   \alpha_{zz} = \sum_{k\ne 0}^{\infty }  \frac{ \langle \Psi_0 \vert \hat{z} \vert \Psi_k \rangle 
                                      \langle \Psi_k \vert \hat{z}  \vert \Psi_0 \rangle }
                                     { E_k - E_0} 
            \label{eq:SOS}
\end{equation}
Here, $ E_0 $ is the ground-state electronic energy and $ E_k $ is the energy of the k-th excited state and
and $\Psi_0$ is the ground-state wavefunction and
 $\Psi_k$ is the wavefunction of the k-th excited state.
The matrix element $\langle \Psi_0 \vert \hat{z} \vert \Psi_k\rangle$ is the $z^{th}$ component of the 
transition dipole moment. Thus, 
knowledge of the excitation energies and the transition dipole moments can be used to 
compute the polarizability using the SOS method. The summation in the SOS expression (Eq. (\ref{eq:SOS}))
is over all excited states. 
One major advantage of the SOS method is that the transitions that make significant 
contributions to the polarizability can be identified.  If the perturbation is small then one expects the
largest term in this expression to be the first because it has the smallest energy denominator.
In mean-field theories this energy difference
is often approximated by the difference between the energy of highest-occupied molecular orbital (HOMO)
and the energy of the lowest unoccupied molecular orbital (LUMO).  In the following we call the exact first
energy difference, $E_1 - E_0$, and approximations to it the HOMO-LUMO gap.

Exact polarizibilities can also be obtained
using small finite fields.  In the finite field (FF) method the 
molecular Schr\"odinger equation is solved  for different directions
and magnitudes
of the applied electric field
and finite differences
are used to calculate the components, $\alpha_{ij}$, 
of the polarizability tensor. The polarizability matrix elements can be obtained 
using a least square fit to the total molecular energies or by using 
a suitable numerical approximation to  obtain the second derivative. 

The methods for the calculation of the molecular energy by solving Schr\"odinger equation 
can be broadly sorted in two classes: the traditional quantum mechanical methods such as the
HF method and beyond,
and and models based on density functional theory (DFT). In this article, we perform a 
comparative study of polarizabilities computed using the SOS and finite-field 
methods.  The comparison is carried out for polarizabilities calculated within 
the HF approximation and two DFT models.

     Within mean-field theories such as 
the HF or DFT, the SOS equation \ref{eq:SOS} can be written as 
\begin{equation}
\alpha_{ij} = 2 \sum^{unocc.}_{m} \sum^{occ.}_{n}  
      \frac{<\psi_m|p_i|\psi_n><\psi_n|p_j|\psi_m> }{  \epsilon_m - \epsilon_n  }. 
\label{eq:sos}
\end{equation}
Here, the $\{\psi\}$ and \{$\epsilon$\} are the self-consistent molecular orbitals and eigenvalues, respectively.
The summation in this expression runs over all occupied and virtual states. In practice, however, 
only a finite  number of virtual states are included since the contributions from very high energy states are negligible. 
Hereafter, we shall refer to the  Eq. (\ref{eq:sos})
as the {\em sum-over-states} SOS expression  (and not Eq. (\ref{eq:SOS}), which is over fully correlated electronic states )
and the polarizability calculated using this equation as the \afsos.

The single-particle eigenvalues in Eq. (\ref{eq:sos}) can be viewed as
approximations to the excitation energies
of the computational model. 
This expression treats each orbital as responding independently
to the applied electric field, but as the orbitals respond to the applied field the
self-consistent or mean-field component of the Fock and KS potentials also changes\cite{Mahan80}.
Eq. (\ref{eq:sos}) uses only the unperturbed eigenfunctions
{$\psi_m$} and eigenvalues. The additional field-induced electron 
redistribution, or {\em screening}, is missing in the 
computation of polarizabilities by the SOS method. The \afsos\, are called the
unscreened polarizability.   

 The self-consistent 
treatment of field-induced polarization  that  includes the screening effects  
can be computed by the FF method. In 
this method a term $(-\vec{F}\cdot\vec{r})$ that represents interaction between the electrons
and the applied field is added to the zero-field, one-body Hamiltonian, and the single-particle 
equations (Hartree-Fock or Kohn-Sham) are solved self-consistently.  The self-consistent 
procedure takes into account the screening effects or field-induced polarizations.
The polarizability is then calculated by numerical differentiation of 
the energy or dipole moment obtained at various field strengths in limit of 
vanishing field\cite{FF}.
As a result of the complete description of the first-order screening effects, the 
\afFF\, polarizabilities generally are found to 
be in good agreement with experimental values. 
Equivalently, the
energy derivative required for the polarizability can be computed analytically by differentiation 
of the field-dependent Hamiltonian\cite{CHF,CPHF}. The latter approach is called {\em coupled} HF 
method.

                              The lack of an appropriate description of the 
screening  effects in the \afsos calculations generally results in \afsos 
overestimating the experimental polarizabilities and \afFF in density-functional methods.
Early calculations of Stott and Zaremba 
showed that the inclusion of screening by self-consistent calculation of polarizabilities of atoms
reduces its value by roughly 40\%\cite{Zaremba}. Later, Mahan also noted that a self-consistent 
treatment reduced the linear polarizability by about 40\% but the next two higher (hyper) polarizabilities 
are reduced by $2\%$ only\cite{Mahan80}. 
 Similar conclusions were drawn by Pederson and Quong, who  in their calculation 
on the $C_{60}$ fullerene found \afsos to be roughly  three times larger 
than the screened \afFF,
with the latter value being in good agreement with experimental measurements\cite{C60}.
These works show that
the inclusion of screening effects is necessary for a satisfactory estimation 
of polarizability.  The appropriate treatment of screening effects using 
finite-field or equivalent methods requires  full self-consistent solution 
cannot take advantage of the full unperturbed point-group symmetry.   
This lack of symmetry can make polarizability calculations
computationally intensive for large systems such as carbon fullerenes, 
quantum dots, finite nanotubes  etc.  For this reason,  simple schemes to 
correct \afsos\, for the screening effects have been devised and applied\cite{Bertsch,Louie,Pacheco}.
Recently, Gueorguiev, Pacheco and  Tom\'anek have used  a similar scheme to  calculate polarizabilities of 
large carbon fullerenes within the tight-binding method.\cite{Pacheco}. 

 The orbital eigenvalues that are required 
in the SOS method  (Eq. (\ref{eq:sos})) have different interpretations in HF theory and 
DFT \cite{DFT}.
In HF theory, Koopmans theorem states that the occupied orbital eigenvalues are 
the negative of unrelaxed ionization energies.\cite{Koopmans}. 
On the other hand, in DFT only the highest occupied eigenvalue has
a  physical interpretation as the negative of the first ionization potential\cite{ehomo}.
The DFT eigenvalues are the derivatives of 
the total energy with respect to orbital occupancies\cite{Janak}. 
 The practical DFT applications require approximations to the exchange-correlation 
functionals that model the exchange-correlations effects that are quantal in origin.
 Due to the nature of most of these approximate functional forms, the 
effective potential that an electron experiences is contaminated with self-interaction.
Consequently, the negative of the energy eigenvalue of the highest occupied orbital 
in these approximate schemes often underestimates the first ionization potential.
The HOMO-LUMO gap is also underestimated
typically by about 30-40\%\cite{GODBY,Fleszar} due to the so called 
band gap problem\cite{SHAM,Jones}.  In the HF theory, self-Coulomb potential
exactly cancels the self-exchange potential for the occupied electrons and hence 
in the HF model the occupied electrons experience only electronic interactions with the
other occupied electrons.  Thus HF is self-interaction 
free for occupied orbitals. The virtual orbitals, however, experience the full
electronic interaction with all the occupied electrons.  Therefore, the HOMO-LUMO gap in HF theory 
is usually overestimated.

  The unscreened SOS method  (Eq. \ref{eq:sos}) is often used within the HF 
for qualitative purposes.  A better and more appropriate expression in place of 
the HF eigenvalue differences in the denominator  can be obtained based on the 
following argument. Once the zero-field ground-state HF determinant is found, it 
is possible to develop a multiconfigurational treatment which is successively more
accurate by first constructing one-electron excitations, then two-electron
excitations etc. Brillouin's theorem tells us that there is no coupling
between one-electron excitations and the occupied, zero-field molecular orbitals. When
the many-electron field-induced perturbation ($\sum_i \vec{F}  \cdot \vec{r}_i $)
is added to this Hamiltonian, the only coupling between different 
zero-field single-electron excitations is due to this term and it 
is of the form $(\psi_m|\vec{F} \cdot \vec{r}|\psi_n)$ for the excitation
which replaces the formerly occupied state $\psi_m$ by $\psi_n$. The     
energy of this state relative to the ground state is 
$\epsilon_m-\epsilon_n-\Delta_{mn}$ with $\Delta_{mn}$ a gap reducing correction
given by:
\begin{equation}
\Delta_{mn}=(\psi_m \psi_m|1/r_{12}|\psi_n \psi_n)-(\psi_m \psi_n|1/r_{12}|\psi_m 
\psi_n)
\end{equation}
In the above expression the $\psi_m$ are to be viewed as a product state 
of the spatial wavefunction and spinor.  The second term is zero for 
spin-flip excitations.
So, if a CI approach, with only single excitations, is employed, it is
easy to show from perturbation theory that the correct expression for
the polarizability within this approximation is not identical to the
normal sum of states method because there is a pair-dependent 
correction to the energy denominators. 
In other words, rather than Eq. (\ref{eq:sos}), the \afSOS\, should be
\begin{equation}
\alpha_{ij} = 2 \sum^{unocc.}_{m} \sum^{occ.}_{n}
\frac{<\psi_m|p_i|\psi_n><\psi_n|p_j|\psi_m> }{  \epsilon_m - \epsilon_n - \Delta_{mn} }. 
\end{equation}
While the single-electron CI approach is
itself an approximation, it is a more proper definition for the 
unscreened approximation and
differs from the standard SOS method used in HF.

 As the HOMO-LUMO gap is the lowest of many {\em excitation} energies that contribute to \afSOS,  
the trend in  \afSOS\ 
in two types of models (HF and DFT) should be, in general, opposite to that observed for 
the HOMO-LUMO gaps.  While the general 
trends of overestimation and underestimation of \afSOS in 
the HF method and DF based models are to be expected, a detailed
comparison on a large set of molecules is necessary to determine if this trend 
in the two types of models always holds, which partially motivates
the present work.  

Our interest in the SOS method (Eq. \ref{eq:sos})
is primarily due to the fact that it is computationally inexpensive to apply,
particularly for 
large, symmetric molecules for which the  application of  the finite-field method becomes very expensive.
The SOS (using Eq. (\ref{eq:sos})) indeed has been employed in the past to compute the polarizability 
of carbon fullerenes\cite{Bertsch,B_C60,Louie,Pacheco}. In these works, the calculated \afSOS\, were scaled 
using a scaling factor to account for the screening effects.
As the point group symmetry of molecules can be efficiently employed in \afSOS calculations,
Eq. (\ref{eq:sos}) offers a possibility  of obtaining 
estimates of polarizability of large symmetric molecules that are beyond the reach of the more
accurate, but symmetry-breaking, FF method.  Our goal is to estimate the polarizabilities of carbon 
fullerenes, quantum dots, and nanocrystalline materials containing about 1000-2000 atoms
by the SOS method \cite{Big}. The present work is a detailed
investigation to understand 
the behavior of \afSOS\, within different computational schemes.

We have chosen 
a large set of  142  molecules belonging to the extended G2 set of molecules 
whose geometries are well known.
To facilitate the comparison between different 
single-particle methods, we choose the same set of geometries  and basis sets in 
our polarizability calculations in all models. This eliminates some
sources of discrepancy in the comparison of polarizabilities. The 
geometries of 142 molecules were first optimized  
using the B3LYP hybrid DFT model using the 6-311G** orbital basis set.\cite{basis}
B3LYP is one of the most popular 
hybrid functionals. It contains  a weighted mixture of 
the Becke88 exchange functional, the Lee-Yang-Parr  gradient corrected correlation,
Vosko-Wilk-Nusair correlation functional, the local-density exchange functional, 
and HF exchange\cite{B3LYP_A,B3LYP_B}. The mixing coefficients are determined empirically
by fitting to atomization energies.  The HF and B3LYP calculations were 
performed using the Gaussian03\cite{G03} at the Army Research Laboratory
while the PBE calculations were performed 
using the NRLMOL suites of code. The NRLMOL is a massively parallelized suite of 
codes developed at the Naval Research Laboratory for DFT
calculations for molecules\cite{NRLMOL}. The molecular orbitals in the NRLMOL 
are expressed as a sum of linear combinations of Gaussian orbitals and the 
polarizability is computed using the FF method by numerically differentiating 
the calculated dipole moment at different field strengths/directions. The electric 
field vaule are chosen in the step of 0.005 a.u. These values 
provide reliable estimate of the polarizability.  For a recent review 
on the details of polarizability calculations using the NRLMOL code,
we refer the interested reader to Ref. \onlinecite{Mark_Review}.
 It is well known that for accurate computation of polarizability, well chosen Gaussian 
basis set augmented with diffused functions is often necessary. In this work, 
we have used the 6-311++G(d,p) type of basis for all  polarizability calculations.
This basis was selected on two criteria: (i) it is large enough to 
provide good accuracy for the polarizability (ii) its availability for all elements 
that belong to the chosen set of molecules. 

               The \afSOS\, and \afFF of 142 molecules calculated
within HF, B3LYP, and PBE
models are compared in Table \ref{tab:pol}. These are 
the mean polarizabilities obtained from the trace  of the polarizability 
tensor as $\alpha =  \frac{1}{3} \sum_{i=1}^3 \alpha_{ii}$. 
Although the
main goal of this work is to compare screened and unscreened polarizabilities, for completeness
 we also included in the table the experimental values of polarizabilities 
for selected molecules.
 All experimental values are from the CRC Handbook\cite{CRC}. 
The overall agreement between theory and experiment is good. There are 
large differences for some molecules (e.g. ClF$_3$ which has a large 
vibrational contribution to polarizability). There are many possible 
causes for these differences such as temperature effects, basis sets effects 
and vibrational contributions. Proper treatment of these effects will give 
better agreement but is outside the scope and purpose of  the
present paper.  Hence we shall not elaborate on 
the comparison of theoretical and experimental values of polarizability.
The comparison of the (unscreened)  \afSOS\,  with 
screened  \afFF, within the three models show 
different trends.  In the HF method the \afSOS\, are smaller than the \afFF 
\, with the exceptions of the  P$_2$, CS, COS, and N$_2$O molecules.
The two values of polarizability for these exceptions are quite close, with
the difference for  P$_2$ being the largest.
This almost always correct trend of HF polarizabilities is reversed in the
parameter-free PBE DFT model.  We find no
exceptions.  In PBE, SOS polarizabilities are always less than FF polarizabilities.
The \afSOS\, in PBE are consistently higher than the  \afFF 
\, polarizabilities. These trends, as noted earlier, are consistent with  the
general trend of the HF and PBE models to underestimate and overestimate
the HOMO-LUMO energy gap. The  \afSOS\, are also predicted to be larger than \afFF 
\, in the B3LYP method. However, the overestimation in the B3LYP model
is less than that in the PBE model. 
 The mean absolute error (MAE) 
calculated as 
$MAE = \frac{1}{142}\sum_{i=1}^{142}\vert\alpha_{SOS}-\alpha_{FF}\vert $ is 
reduced from 3.55 \AA$^3$ for the PBE to 1.95 \AA$^3$ for the B3LYP functional.  The MAE is least 
in the HF theory, for which the mean error is negative. The hybrid B3LYP functional
benefits from the opposite effects 
of overestimation in pure DF models and underestimation in the HF theory.

%--------------------------------------------------------

\LTchunksize=270
\begingroup
\squeezetable
\begin{longtable*}{lccccccc}
\caption{\label{tab:pol}The polarizabilities  for the set of 142 (extended G2 set) of 
molecules computed by sum over states and coupled HF methods or finite field
within the Hartree-Fock approximation, hybrid B3LYP model and the PBE generalized
gradient approximation. The polarizabilities are calculated using 
the 6-311++G(d,p) at the geometries optimized at B3LYP/6-311G** level.
All values are in \AA$^3$. I: SOS , II:  Coupled HF or finite-field}
\\
\hline\hline
 Molecule &       HF(I) & HF(II) & B3LYP(I) & B3LYP (II) & PBE (I) & PBE (II) & Expt.\cite{CRC}\\
\hline 
\hline
       H$_2$&   0.3&   0.4&   0.5&   0.4&   0.6&   0.5 &   0.8 \\ 
         LiH&   1.9&   3.2&   4.0&   4.2&   5.5&   4.7 &   \\ 
         BeH&   2.2&   4.2&   4.4&   4.5&   3.2&   4.7 &   \\ 
          CH&   1.1&   1.7&   2.0&   1.8&   2.6&   1.8 &   \\ 
 CH$_2$($^3B_1$)&   1.1&   1.5&   1.8&   1.6&   1.6&   1.7 &   \\ 
 CH$_2$($^1A_1$)&   1.4&   1.9&   2.5&   2.1&   3.7&   2.1 &   \\ 
      CH$_3$&   1.4&   1.8&   2.2&   2.0&   2.2&   2.0 &   \\ 
      CH$_4$&   1.7&   2.1&   2.6&   2.2&   3.0&   2.2 &   2.6 \\ 
          NH&   0.7&   1.1&   1.2&   1.1&   1.5&   1.2 &   \\ 
      NH$_2$&   1.0&   1.3&   1.6&   1.4&   2.0&   1.5 &   \\ 
      NH$_3$&   1.3&   1.6&   2.1&   1.8&   2.4&   1.8 &   2.8 \\ 
          OH&   0.5&   0.7&   0.8&   0.8&   1.0&   0.8 &   \\ 
      H$_2$O&   0.8&   1.0&   1.2&   1.1&   1.5&   1.1 &   \\ 
          HF&   0.4&   0.5&   0.6&   0.5&   0.7&   0.6 &   \\ 
      Li$_2$&  17.7&  30.3&  37.9&  29.3&  57.0&  30.2 &  32.8 \\ 
         LiF&   0.6&   0.8&   1.3&   1.2&   1.7&   1.4 &   \\ 
  C$_2$H$_2$&   2.6&   2.8&   4.3&   2.9&   5.1&   2.9 &   \\ 
  C$_2$H$_4$&   3.1&   3.8&   5.1&   3.8&   6.1&   3.8 &   \\ 
  C$_2$H$_6$&   3.1&   3.8&   5.0&   4.0&   5.9&   4.1 &   \\ 
          CN&   1.6&   1.9&   4.3&   2.9&   5.9&   2.7 &   \\ 
         HCN&   2.0&   2.1&   3.3&   2.1&   3.9&   2.2 &   \\ 
          CO&   1.4&   1.6&   2.5&   1.7&   3.0&   1.8 &   1.9 \\ 
         HCO&   1.7&   2.1&   3.0&   2.2&   3.7&   2.3 &   \\ 
 H$_2$CO (formaldehyde)&   1.8&   2.2&   3.1&   2.3&   3.8&   2.4 &   2.8 \\ 
 H$_3$CO$_{}$H &   2.1&   2.6&   3.4&   2.8&   4.1&   2.9 &   \\ 
       N$_2$&   1.4&   1.5&   2.4&   1.5&   2.8&   1.6 &   1.7 \\ 
  N$_2$H$_4$&   2.3&   2.8&   3.8&   3.1&   4.5&   3.2 &   \\ 
       O$_2$&   1.1&   1.4&   1.8&   1.2&   3.2&   1.2 &   1.6 \\ 
  H$_2$O$_2$&   1.4&   1.8&   2.4&   1.9&   2.9&   1.9 &   \\ 
       F$_2$&   0.7&   1.0&   1.3&   0.9&   1.6&   0.9 &   \\ 
      CO$_2$&   1.9&   2.0&   3.4&   2.2&   4.2&   2.2 &   2.9 \\ 
 SiH$_2$($^1A_1$)&   3.2&   4.4&   6.0&   4.5&   7.9&   4.6 &   \\ 
 SiH$_2$($^3B_1$)&   2.6&   3.7&   4.7&   4.0&   4.3&   4.1 &   \\ 
     SiH$_3$&   2.9&   3.9&   5.1&   4.2&   5.0&   4.3 &   \\ 
     SiH$_4$&   2.9&   3.7&   5.0&   4.0&   6.2&   4.2 &   5.4 \\ 
      PH$_2$&   2.5&   3.3&   4.4&   3.4&   5.3&   3.5 &   \\ 
      PH$_3$&   2.9&   3.6&   5.0&   3.7&   6.0&   3.7 &   4.8 \\ 
      H$_2$S&   2.2&   2.6&   3.6&   2.8&   4.3&   2.8 &   4.0 \\ 
         HCl&   1.3&   1.5&   2.0&   1.6&   2.3&   1.6 &   2.8 \\ 
      Na$_2$&  24.8&  40.2&  48.9&  34.0&  70.4&  35.4 &   \\ 
      Si$_2$&   9.2&  13.8&  18.4&  14.6&  26.1&  19.1 &   \\ 
       P$_2$&   7.1&   6.3&  13.1&   6.2&  16.7&   6.2 &   \\ 
       S$_2$&   4.6&   5.5&   8.4&   4.8&  15.6&   4.7 &   \\ 
      Cl$_2$&   2.8&   3.3&   4.7&   3.2&   5.7&   3.2 &   4.6 \\ 
        NaCl&   2.2&   2.9&   4.5&   4.0&   6.0&   4.4 &   \\ 
         SiO&   3.2&   3.7&   6.0&   3.9&   7.7&   4.0 &   \\ 
          CS&   3.5&   3.4&   6.4&   3.6&   8.1&   3.6 &   \\ 
          SO&   2.3&   2.8&   4.1&   2.7&   7.0&   2.7 &   \\ 
         ClO&   1.7&   2.0&   3.5&   2.3&   4.5&   2.3 &   \\ 
         ClF&   1.5&   1.8&   2.6&   1.9&   3.2&   1.9 &   \\ 
 Si$_2$H$_6$&   6.3&   7.8&  11.4&   8.5&  14.2&   8.9 &   \\ 
   CH$_3$Cl &   2.9&   3.4&   4.6&   3.6&   5.4&   3.6 &   5.3 \\ 
 H$_3$CSH   &   3.9&   4.5&   6.4&   4.8&   7.6&   4.8 &   \\ 
        HOCl&   2.0&   2.5&   3.4&   2.5&   4.2&   2.5 &   \\ 
      SO$_2$&   3.1&   3.2&   6.1&   3.4&   7.9&   3.5 &   3.7 \\ 
      BF$_3$&   1.4&   1.7&   2.4&   2.0&   3.0&   2.2 &   3.3 \\ 
     BCl$_3$&   5.7&   6.2&   9.9&   7.0&  12.2&   7.2 &   9.4 \\ 
     AlF$_3$&   1.6&   2.0&   3.0&   2.6&   3.8&   2.9 &   \\ 
    AlCl$_3$&   6.0&   6.9&  10.7&   8.0&  13.1&   8.3 &   \\ 
      CF$_4$&   1.8&   2.1&   3.0&   2.5&   3.6&   2.6 &   3.8 \\ 
     CCl$_4$&   7.6&   8.4&  13.4&   9.1&  16.5&   9.4 &  11.2 \\ 
         COS&   4.1&   4.0&   7.7&   4.2&   9.7&   4.3 &   \\ 
      CS$_2$&   7.4&   6.9&  14.7&   7.0&  19.2&   7.0 &   8.7 \\ 
     COF$_2$&   1.8&   2.1&   3.3&   2.4&   4.0&   2.5 &   \\ 
     SiF$_4$&   1.9&   2.3&   3.4&   2.9&   4.2&   3.1 &   5.5 \\ 
    SiCl$_4$&   7.9&   8.7&  13.9&   9.9&  17.0&  10.2 &   \\ 
      N$_2$O&   2.5&   2.4&   4.8&   2.5&   6.1&   2.6 &   3.0 \\ 
 C$_2$Cl$_4$&   8.8&  10.0&  16.1&  10.8&  20.2&  11.1 &   \\ 
    CF$_3$CN&   3.5&   3.8&   6.1&   4.3&   7.6&   4.5 &   \\ 
 CH$_3$CCH (propyne)&   4.1&   4.6&   6.9&   4.9&   8.4&   5.0 &   6.2 \\ 
 CH$_2$CCH$_2$ (allene)&   4.6&   5.6&   8.0&   5.6&   9.9&   5.6 &   \\ 
 C$_3$H$_4$ (cyclopropyne)&   4.3&   4.8&   7.0&   4.9&   8.4&   5.0 &   6.2 \\ 
 CH$_3$CHCH$_2$ (propene)&   4.7&   5.5&   7.7&   5.7&   9.3&   5.8 &   \\ 
  C$_3$H$_6$&   4.4&   5.0&   7.1&   5.3&   8.3&   5.4 &   \\ 
 C$_3$H$_8$ (propane)&   4.7&   5.5&   7.5&   5.8&   8.9&   6.0 &   6.3 \\ 
 CH$_2$CHCHCH$_2$ (butidene)&   6.6&   8.0&  11.9&   8.0&  15.1&   8.1 &   \\ 
 C$_4$H$_6$ (butyne)&   5.6&   6.5&   9.6&   7.0&  11.7&   7.2 &   \\ 
 C$_4$H$_6$ (methylene cylcopropane)&   5.9&   7.0&  10.0&   7.2&  12.1&   7.3 &   \\ 
 C$_4$H$_6$ (bicyclobutane)&   5.7&   6.2&   9.3&   6.5&  11.0&   6.7 &   \\ 
 C$_4$H$_6$ (cyclobutene)&   5.8&   6.5&   9.6&   6.8&  11.6&   6.9 &   \\ 
 C$_4$H$_8$ (Cyclobutane)&   5.8&   6.6&   9.3&   7.0&  11.0&   7.2 &   \\ 
 C$_4$H$_8$ (isobutene)&   6.2&   7.2&  10.4&   7.6&  12.5&   7.8 &   \\ 
 C$_4$H$_{10}$ (butane)&   6.2&   7.2&  10.0&   7.7&  11.8&   7.9 &   8.2 \\ 
 C$_4$H$_{10}$ (isobutane)&   6.2&   7.2&  10.1&   7.7&  12.0&   7.9 &   8.1 \\ 
 C$_5$H$_{8}$ (spiropentane)&   7.1&   8.1&  11.7&   8.5&  13.9&   8.8 &   \\ 
 C$_6$H$_{6}$ (benzene)&   9.4&   9.6&  15.6&   9.8&  18.9&   9.9 &  10.0 \\ 
 CH$_2$F$_{2}$ (difluromethylene)&   1.6&   2.0&   2.7&   2.3&   3.3&   2.4 &   \\ 
 CHF$_3$ (trifluromethane)&   1.7&   2.1&   2.9&   2.4&   3.5&   2.5 &   3.5 \\ 
 CH$_2$Cl$_{2}$ (dichloromethane)&   4.3&   5.0&   7.2&   5.3&   8.7&   5.5 &   6.5 \\ 
 CHCl$_3$ (chloroform)&   5.9&   6.7&  10.2&   7.2&  12.5&   7.5 &   9.5 \\ 
 CH$_3$NH$_2$ (methylamine)&   2.7&   3.2&   4.3&   3.5&   5.1&   3.6 &   \\ 
 CH$_3$CN (methyl cynaide)&   3.4&   3.8&   5.7&   4.0&   6.9&   4.1 &   \\ 
 CH$_3$NO$_{2}$ (nitromethane)&   3.7&   4.2&   6.9&   4.5&   8.7&   4.6 &   7.4 \\ 
 CH$_3$ONO (methyl nitrite)&   3.6&   4.0&   6.7&   4.5&   8.5&   4.7 &   \\ 
 CH$_3$SiH$_{3}$ (methyl silane)&   4.5&   5.5&   7.6&   5.9&   9.3&   6.2 &   \\ 
 CHOOH (formic acid)&   2.3&   2.6&   4.1&   3.0&   5.2&   3.2 &   3.4 \\ 
 HCOOCH$_3$ (methyl formate)&   3.7&   4.3&   6.4&   4.7&   7.8&   4.9 &   5.0 \\ 
 CH$_3$CONH$_{2}$ (acetamide)&   4.2&   4.9&   7.4&   5.5&   9.2&   5.8 &   5.7 \\ 
 C$_2$H$_4$NH (aziridine)&   3.9&   4.4&   6.3&   4.7&   7.5&   4.8 &   2.6 \\ 
 CNCN   (cyanogen)&   4.1&   4.2&   8.0&   4.4&  10.4&   4.5 &   \\ 
 (CH$_3$)$_{2}$NH (dimethylamine)&   4.2&   5.0&   6.9&   5.4&   8.3&   5.7 &   \\ 
 CH$_3$CH$_2$NH$_2$ (trans ethyalmine)&   4.2&   5.0&   7.0&   5.5&   8.4&   5.7 &   \\ 
 CH$_2$CO (ketene)&   3.2&   3.6&   5.8&   3.8&   7.1&   3.9 &   \\ 
 C$_2$H$_{4}$O (oxirane)&   3.2&   3.7&   5.3&   4.0&   6.4&   4.1 &   \\ 
 CH$_3$CHO   (acetaldehyde)&   3.3&   3.9&   5.7&   4.3&   7.0&   4.5 &   \\ 
 HCOCOH (glyoxal)&   3.5&   4.0&   6.3&   4.4&   8.0&   4.6 &   \\ 
 CH$_3$CH$_2$OH (ethanol)&   3.6&   4.3&   5.9&   4.7&   7.0&   4.8 &   \\ 
 (CH$_3$)$_{2}$O (dimethylether)&   3.6&   4.3&   5.9&   4.7&   7.1&   4.9 &   \\ 
 C$_2$H$_4$S (thioxirane)&   5.4&   5.9&   9.1&   6.1&  10.9&   6.2 &   \\ 
 (CH$_3$)$_2$SO (dimethyl sulfoxide)&   6.1&   6.9&  10.9&   7.6&  13.5&   7.9 &   \\ 
 CH$_3$CH$_2$SH (ethanethiol)&   5.5&   6.4&   9.1&   6.7&  10.8&   6.9 &   \\ 
 (CH$_3$)$_2$S (dimethyl sulphide)&   5.5&   6.4&   9.2&   6.8&  11.0&   6.9 &   \\ 
 CH$_2$CHF (vinyl fluride)&   3.0&   3.6&   5.2&   3.8&   6.3&   3.9 &   \\ 
 CH$_3$CH$_2$Cl (ethyl chloride)&   4.5&   5.3&   7.4&   5.6&   8.8&   5.7 &   \\ 
 CH$_2$CHCl (vinyl chloride)&   4.4&   5.2&   7.6&   5.4&   9.3&   5.5 &   \\ 
 CH$_3$CHCN  (acrylonitrile)&   5.1&   5.8&   9.0&   5.9&  11.4&   6.0 &   \\ 
 (CH$_3$)$_2$CO (acetone)&   4.8&   5.5&   8.1&   6.0&   9.9&   6.3 &   6.3 \\ 
 CH$_3$COOH (acetic acid)&   3.7&   4.3&   6.4&   4.7&   7.8&   5.0 &   5.1 \\ 
 CH$_3$COF (acetyl fluride)&   3.2&   3.7&   5.5&   4.1&   6.8&   4.3 &   \\ 
 CH$_3$COCl (acetyl chloride)&   4.9&   5.5&   8.7&   6.0&  10.8&   6.2 &   \\ 
 CH$_3$CH$_2$CH$_2$Cl (propyl chloride)&   6.0&   7.0&   9.9&   7.5&  11.8&   7.7 &   \\ 
 (CH$_3$)$_2$CHOH (isopropanol)&   5.1&   6.0&   8.5&   6.5&  10.2&   6.8 &   \\ 
 CH$_3$CH$_2$OCH$_3$ (methyl ethylether)&   5.0&   6.0&   8.3&   6.6&  10.0&   6.8 &   \\ 
 (CH$_3$)$_3$N (trimethylamine)&   5.8&   6.7&   9.7&   7.4&  11.8&   7.8 &   \\ 
 C$_4$H$_4$O (furan)&   6.2&   6.6&  10.5&   6.8&  12.7&   7.0 &   \\ 
 C$_4$H$_4$S (thiophene)&   8.6&   8.7&  14.8&   9.0&  18.1&   9.1 &   9.7 \\ 
 C$_4$H$_4$NH (pyrole)&   7.0&   7.4&  11.6&   7.7&  14.0&   7.8 &   \\ 
 C$_5$H$_5$N (pyridine)&   8.6&   8.7&  14.5&   9.1&  17.6&   9.2 &   9.5 \\ 
 CCH  (ethynyl radical)&   2.7&   3.3&   4.9&   3.6&   5.9&   3.7 &   \\ 
    CH$_3$CO&   3.1&   3.8&   5.6&   4.2&   6.6&   4.5 &   \\ 
 CH$_2$OH (hydroxymethyl)&   1.9&   2.4&   3.3&   2.8&   4.1&   2.9 &   \\ 
        ClNO&   3.9&   4.4&   8.7&   4.3&  12.6&   4.4 &   \\ 
      NF$_3$&   1.8&   2.2&   3.3&   2.4&   4.3&   2.5 &   3.6 \\ 
      PF$_3$&   2.5&   3.0&   4.6&   3.4&   5.8&   3.6 &   \\ 
       O$_3$&   2.5&   2.7&   5.5&   2.4&   7.8&   2.4 &   3.2 \\ 
      F$_2$O&   1.4&   1.8&   2.6&   1.8&   3.4&   1.8 &   \\ 
     ClF$_3$&   3.0&   3.7&   6.4&   3.8&   8.8&   3.9 &   6.3 \\ 
  C$_2$F$_4$&   3.0&   3.5&   5.4&   3.9&   6.9&   4.1 &   \\ 
 CH$_3$O (methoxy radical)&   1.9&   2.4&   3.3&   2.7&   4.2&   2.9 &   \\ 
 CH$_3$CH$_2$O&   3.4&   4.1&   5.8&   4.6&   7.3&   4.8 &   \\ 
 CH$_3$S (methylsulfide radical)&   3.3&   4.0&   5.5&   4.3&   6.7&   4.4 &   \\ 
 CH$_3$CH$_2$ (ethyl radical)&   2.9&   3.5&   4.7&   3.9&   5.6&   4.0 &   \\ 
\hline
\end{longtable*}
\endgroup

%--------------------------------------------------------

   As mentioned earlier, empirical schemes that convert the SOS polarizabilities 
to screened polarizabilities have been devised and applied to carbon 
fullerenes\cite{Bertsch,Louie,B_C60,Tomanek}. These schemes were usually 
obtained in the random phase approximation and were applied 
to spherical molecules like fullerenes and cylindrical systems  
like nanotubes.  Their application to molecules with arbitrary 
geometry has not yet been reported.
An alternative possibility to improve upon the SOS polarizability will be to 
improve the approximation for the exchange-correlation functional.
Correcting for the self-interaction in the approximate functional 
may improve the SOS polarizabilities. It has been found that the  
self-interaction corrections (SIC) significantly improves the eigenvalues 
and the band gap\cite{Perdew}.  The SIC implementation  is however 
quite complicated to implement and computationally expensive\cite{Mark_SIC}.
The time dependent DFT (TD-DFT) within the linear response 
regime is another popular method for obtaining excitation energies but like 
the finite-field method it breaks the point-group symmetry.
Much simpler approach would be to assume 
that the orbitals or the transition dipole moments obtained in the 
approximate DFT models to be reasonably accurate and correct only 
the eigenvalues using simple schemes.

 There are a number of methods in the literature that try to improve upon the 
HOMO-LUMO energy gap in the approximate DFT models typically by correcting the 
eigenvalues of the single-particle equations. Such corrections should 
result in some improvements in the SOS polarizability.  Most of these methods
are, however, quite complex and computationally expensive, limiting 
their advantages for correcting \afSOS\
over the finite field method in terms of 
computational time.
Hence simple schemes for correcting the eigenvalues are more appealing\cite{Jellinek,Trickey,Ballone} 
and worth exploring in calculation of \afSOS.  Such simple schemes have 
been used in calculations of photoelectron spectra with some success\cite{PES}.
The simplest such a correction would to be replace the \DHL\, by 
the quasiparticle gap.  The latter set of quantities can be computed by finite 
difference method or the so called $\Delta SCF$ method. This 
requires two additional self-consistent calculations, one for 
cation  and other for anion. 
The correction $\delta$ to the \DHL\, is then 
$E(N-1)+E(N+1)-2E(N)-\epsilon_{HOMO}+\epsilon_{LUMO}$, where, $E(N)$ is the 
self-consistent total energy of the system containing $N$ electrons, 
$\epsilon_{HOMO/LUMO}$  are the eigenvalues of the HOMO/LUMO of the neutral molecule.
This correction, when applied to all single particle energies, 
would shift the occupied and unoccupied eigenvalues in opposite directions 
and hence may be useful in getting better estimate of \afSOS.
Its application to the $Li_2$ for which the difference 
between the \afFF\, and \afSOS\, is large, indicate that this 
correction overcorrects \afsos\,, roughly by a factor of two
(from  57.0 A$^3$ to 19.0 \AA$^3$ for PBE).

 Another possible scheme can be devised by  noting that the HF 
overestimates the \afSOS\, whereas the PBE 
underestimates.  It is possible  to mix  
the HF \afSOS\, and the PBE  \afSOS\, to obtain finite-field quality 
polarizability.  In absence of more accurate data, we choose B3LYP 
finite-field polarizabilities as a target set.
The following interpolation scheme can be used:
$\alpha^{mix} = (1-x)*\alpha^{SOS}_{HF} + x \alpha^{SOS}_{PBE} $, where the parameter
 $x\, \, (0 < x < 1)$  can be determined  by 
minimizing the mean absolute error (MAE) in the $\alpha^{mix}$ and $\alpha^{FF}_{B3LYP}$.
The minimization procedure  gives the optimal mixing parameter to be 0.22. 
The MAE at the minimum is  0.4 \AA$^3$. 
Thus about 80\% of HF \afSOS\, mixed with about 20\% of 
the PBE \afSOS\, will give \afSOS\, comparable to the B3LYP \afFF.    
In this
application more HF than DFT is required, whereas Becke needed less HF than DFT for optimizing B3LYP for
atomization energies\cite{B3LYP_A}.
This procedure requires calculation of \afSOS\, in two models 
which could be performed efficiently by making use of any symmetry that system may possess.
Thus it could be applied to symmetric quantum dots, fullerenes and nanocrystals to obtain estimates 
of polarizability. 
We note that similar idea was explored for a few molecules by Dunlap and Karna\cite{Dunlap}.
These authors
used average values of eigenvalues in the HF and the local-density approximation 
 (DFT) eigenvalues in the SOS expression (Eq. \ref{eq:sos}).

 In conclusion, a systematic comparison of the SOS and FF methods for the calculation of
molecular dipole polarizability for a set of 142 molecules from the extended G2 set
has been performed.
The trends in the two sets, \afSOS \,  and \afFF, of polarizabilities are examined in three 
widely used single-particle methods: The HF approximation, the PBE-GGA within
DFT, and the hybrid B3LYP model that mixes DFT exchange with 
HF exchange. In order to minimize the other sources that can lead to differences in 
the polarizabilites in different models, the same set of molecular geometries (optimized 
at the B3LYP/6-311G** level) and  6-311++G(d,p) orbital basis set was used.
The calculations show that \afSOS \, polarizabilities are almost always underestimated 
in the  HF method. However, exceptions to the trend do exist. 
The \afSOS\, of P$_2$, CS, COS, and N$_2$O are overestimated with respect 
to the \afFF\, polarizability.
On the other hand, in the PBE-GGA model, 
\afSOS\, is always overestimated with respect to the \afFF.  These observations correlate 
with the generally observed trend of the respective underestimation and overestimation 
of the HOMO-LUMO gap in the PBE-GGA and HF models. Although the \afSOS \, overestimates the \afFF \,
in the hybrid B3LYP,  the differences in the two polarizabilities is less than 
that observed for the PBE model.   
The comparison of screened polarizabilities  \afFF\, in the three models shows that 
the \afFF\, in B3LYP and PBE models are larger than those calculated 
with the HF method.
Thus inclusion of correlation effects, in general, leads to increase 
in (finite-field) polarizability.
Finally, 
a simple scheme that interpolates \afSOS\, values in HF and PBE
to obtain B3LYP \afFF \, is devised by minimizing the mean absolute error.
A simple scheme like this 
may be useful to estimate polarizabilities of large symmetric molecules such as fullerenes,
quantum dots or nanocrystalline materials.

  Authors acknowledge Dr. S. P. Karna and Prof. P. B. Allen for discussions.
The Gaussian03 calculations were performed at the Army Research Laboratory Major Shared Resource
Center (ARL-MSRC).
This work was supported in part by the Office of Naval Research, directly 
and through the Naval Research Laboratory (ONR Grant No. N000140211046,
and 05PR07548-00) and in parts by the NSF through CREST grant (Grant No. NIRT-0304122), by 
the University of Texas at El Paso.


\begin{thebibliography}{99}
\bibitem{B3LYP_A}
A. D. Becke, J. Chem. Phys.  98, 5648 (1993);

\bibitem{B3LYP_B}
C. Lee, W. Yang, and R. G. Parr, Phys. Rev. B 37, 785 (1988);
B. Miehlich, A. Savin, H. Stoll, and H. Preuss, Chem. Phys. Lett. 157, 200 (1989).


\bibitem{PBE}
  J. P. Perdew, K. Burke and M. Ernzerhof,  Phys.  Rev.  Lett.   {\bf  77 }, 3865(1996).
 
%\bibitem{Born}
%J. H. Jones, Proc. Roy. Soc. (London)  {\bf  105}, 650 (1924);
%M. Born and W. Heisenberg, Z. Phys. {\bf 23}, 388 (1924); L. Pauling, Proc. Roy. Soc.  (London) {\bf 114},
%181(1927).

\bibitem{Reviews}
G. D. Mahan and K. R. Subbaswamy,  {\em Local Density Theory of polarizability}, Plenum Press, New York (1990);
D. M. Bishop, Rev. Mod. Phys. {bf 62}, 343 (1990); 
{\it Electric-Dipole Polarizabilities of Atoms, Molecules and Clusters}, K.D. Bonin and
V.V. Kresin (World Scientific, 1997).

\bibitem{R2}
  For a recent review of this topic, see for example {\em 
 Theoretical and Computational Modelling of NLO and Electronic Materials,}
edited by S.P. Karna and A.T. Yeats (ACS Press, Washington, DC, 1996).



\bibitem{Mahan80}
G. D. Mahan, Phys. Rev. A {\bf 22}, 1780 (1980).


\bibitem{FF}
 H. D. Cohen and C. C. J. Roothan, J. Chem. Phys. 43, S34 (1965);
J. A. Pople, J. W. McIver Jr., and N. S. Ostlund, J. Chem. Phys. {\bf 49}, 2960 (1968);
A. D. Buckingham, Adv. Chem. Phys. {\bf 12}, 107 (1967);
H. A. Kurtz, J. J. P. Stewart, and K. Dieter, J. Comp. Chem. {\bf 11}, 82 (1990).
H. A. Kurtz, I. J. Quant. Chem.  {\bf S24}, 791 (1990);
A. A. Quong and M. R. Pederson, Phys. Rev. B {\bf 46}, 12906 (1992);
J. Guan, M. E. Casida, A. M. K{\"o}ster, and D. R. Salahub, Phys. Rev. B {\bf 52}, 2184 (1995);
S. A. Blundell, C. Guet, and R. R. Zope, Phys. Rev. Lett. {\bf  84},  4826 (2000).

\bibitem{CHF}
C. A. Coulson and H. C. Longuet-Higgins, Proc. Roy. Soc. (London) A {\bf 191}, 39 (1947);
H. Peng, Proc. Roy. Soc. (London), {\bf A178}, 499 (1941); L. C. Allen, Phys. Rev. {\bf 118}, 167 (1960);
A. Dalgarno, Adv. Phys. {\bf 11}, 281 (1962).


\bibitem{CPHF}
J. Gerratt and I. M. Mills, J. Chem. Phys. 49, 1719 (1968);
R. McWeeny, Rev. Mod. Phys. 32, 335 (1960);
R. McWeeny, Phys. Rev. 128, 1028 (1961);
R. M. Stevens, R. M. Pitzer, and W. N. Lipscomb, J. Chem. Phys. 38, 550 (1963);
J. L. Dodds, R. McWeeny, W. T. Raynes, and J. P. Riley, Mol. Phys. 33, 611 (1977);
J. L. Dodds, R. McWeeny, and A. J. Sadlej, Mol. Phys. 34, 1779 (1977);
Y. Osamura, Y. Yamaguchi, and H. F. Schaefer III, J. Chem. Phys. 77, 383 (1982);
P. Pulay, J. Chem. Phys. 78, 5043 (1983);
C. E. Dykstra and P. G. Jasien, Chem. Phys. Lett. 109, 388 (1984). 



\bibitem{Zaremba}
M. J. Stott and E. Zaremba, Phys. Rev. A {\bf 21}, 12 (1980); Erratum: {\bf 22}, 2293 (1980).


\bibitem{C60}
M. R. Pederson and A. A. Quong, Phys. Rev. B {\bf 46}, 13584 (1992).

\bibitem{Szabo}
A. Szabo and N. S. Ostlund, {\it Modern Quantum Chemistry: Introduction to Advanced Electronic Structure Theory, 1st ed.},
 McGraw-Hill, New York, 1989.

\bibitem{Bertsch}
 G. F. Bertsch, A. Bulgac, D. Tom\'anek, and Y. Wang,  Phys. Rev. Lett. {\bf 67}, 2690 (1991).

\bibitem{Louie}
L. X. Benedict, S. G. Louie, and M. L. Cohen, Phys. Rev.  B {\bf 52}, 8541 (1995).

\bibitem{Pacheco}
F. Alasia, R. A. Broglia, H. E. Roman, L. L. Serra, G. Colo, anf J. M. Pacheco,
J. Phys. B {\bf 27}, L643 (1994).

%\bibitem{Ruiz_A}
%S. Iglesias-Groth, A. Ruiz,J. Bretón, and J. M. Gomez Llorente,
%J. Chem. Phys. {\bf 116}, 10648 (2002);
%J. Chem. Phys. {\bf 118}, 7103 (2003).

%\bibitem{Ruiz_B}
% A. Ruiz,J. Bretón, and J. M. Gomez Llorente,
%Phys. Rev. Lett. 94, 105501 (2005);
%S. Iglesias-Groth, The Astrophysical Journal, {\bf 632}, L25 (2005).


\bibitem{B_C60}
E. Westin, A. Rosen, G Te Velde and E. J. Baerends, J. Phys. B {\bf 29}, 5087 (1996).


\bibitem{DFT}
P. Hohenberg and W. Kohn, Phys. Rev. 136, B864 (1964);
W. Kohn and L. J. Sham, Phys. Rev. 140, A1133 (1965). 

%\bibitem{Casida}
%J. Guan, P. Duffy, J. T. Carter, D. P. Chong, K. C. Casida, M. E. Casida, and M. Wrinn, J. Chem. Phys. {\bf 98},
%1753 (1993).

%\bibitem{PS}
%Y. Gao, R. A. Friesner, J. Chem. Phys. {\bf 122}, 104102 (2005) and the references therein.

\bibitem{Koopmans}
 T. Koopmans, Physica (Utrecht) {\bf 1}, 104 (1933).

\bibitem{ehomo}
J. P. Perdew, R. G. Parr, M. Levy, and J. L. Balduz, Jr.  Phys. Rev. Lett. {\bf 49}, 1691  (1982);
M. Levy, J. P. Perdew and V.  Sahni, Phys. Rev. A {\bf 30}, 2745 (1984); 
C. O. Almbladh and U. von Barth, Phys. Rev. B {\bf 31}, 3231 (1985);
L. Kleinman Phys. Rev. B {\bf 56}, 16029 (1997);
J. P. Perdew and M. Levy, Phys. Rev. B {\bf 56}, 16021  (1997);
  L. Kleinman, Phys. Rev. B {\bf 56}, 12 042 (1997);
M. E. Casida, Phys. Rev. B {\bf 59}, 4694 (1999);
M. K. Harbola, Phys. Rev. B {\bf 60}, 4545 (1999).


\bibitem{Janak}
J. F. Janak, Phys. Rev. B  {\bf 18}, 7165 (1978).

\bibitem{GODBY}
R. W. Godby, M. Schl\"uter, and L. J. Sham, Phys. Rev. B {\bf 37}, 10159 (1988).

\bibitem{Fleszar}
A. Fleszar, Phys. Rev. B {\bf 64}, 245204 (2001).

\bibitem{SHAM}
J. P. Perdew and M. Levy, Phys. Rev. Lett. {\bf 51}, 1884 (1983); L. J. Sham and M. Schl\"uter,
{\it ibid} {\bf 51}, 1888 (1983).

\bibitem{Jones}
R. O. Jones and O. Gunnarsson, Rev. Mod. Phys. {\bf 61}, 689 (1989).

\bibitem{Big}
 R. R. Zope and B. I. Dunlap, Chem. Phys. Lett. {\bf 422}, 451 (2006);
 Y. Zhao, Y.-H. Kim, M.-H. Du, and S. B. Zhang, Phys. Rev. Lett. 93, 015502 (2004).


\bibitem{basis}
 A. D. McLean and G. S. Chandler, J. Chem. Phys. {\bf 72}, 5639 (1980);
R. Krishnan, J. S. Binkley, R. Seeger, and J. A. Pople, J. Chem. Phys. {\bf 72}, 650 (1980)




\bibitem{G03}
Gaussian 03, Revision C.02, M. J. Frisch, G. W. Trucks, H. B. Schlegel, G. E. Scuseria, M. A. Robb, J. R. Cheeseman, J. A. Montgomery, Jr., T. Vreven, K. N. Kudin, J. C. Burant, J. M. Millam, S. S. Iyengar, J. Tomasi, V. Barone, B. Mennucci, M. Cossi, G. Scalmani, N. Rega, G. A. Petersson, H. Nakatsuji, M. Hada, M. Ehara, K. Toyota, R. Fukuda, J. Hasegawa, M. Ishida, T. Nakajima, Y. Honda, O. Kitao, H. Nakai, M. Klene, X. Li, J. E. Knox, H. P. Hratchian, J. B. Cross, V. Bakken, C. Adamo, J. Jaramillo, R. Gomperts, R. E. Stratmann, O. Yazyev, A. J. Austin, R. Cammi, C. Pomelli, J. W. Ochterski, P. Y. Ayala, K. Morokuma, G. A. Voth, P. Salvador, J. J. Dannenberg, V. G. Zakrzewski, S. Dapprich, A. D. Daniels, M. C. Strain, O. Farkas, D. K. Malick, A. D. Rabuck, K. Raghavachari, J. B. Foresman, J. V. Ortiz, Q. Cui, A. G. Baboul, S. Clifford, J. Cioslowski, B. B. Stefanov, G. Liu, A. Liashenko, P. Piskorz, I. Komaromi, R. L. Martin, D. J. Fox, T. Keith, M. A. Al-Laham, C. Y. Peng, A. Nanayakkara, M. Challacombe, P. M. W. Gill, B. Johnson, W. Chen, M. W. Wong, C. Gonzalez, and J. A. Pople, Gaussian, Inc., Wallingford CT, 2004.

\bibitem{NRLMOL}
   M. R. Pederson and K. A. Jackson,
           Phys. Rev. B. {\bf 41}, 7453 (1990); {\it ibid} {\bf 43}, 7312 (1991);
    K. A. Jackson and M. R. Pederson, {\it ibid} {\bf 42}, 3276 (1990).


\bibitem{Mark_Review}
 M. R. Pederson and T.Baruah, Lecture series in Computer and Computational Sciences, {\bf 3} 156 (2005). 

\bibitem{CRC}
 C. R. C. Handbook of Chemistry and Physics,  85$th$ edition, (2004-2005).



\bibitem{Tomanek}
G. K. Gueorguiev, J. J. Pacheco, and D. Tom\'anek, Phys. Rev. Lett. {\bf 92}, 215501 (2004).







\bibitem{Perdew}
 J. P. Perdew in {Advances in Quantum Chemistry, ed. S. B. Trickey, Vol. 21, (1990).   }

\bibitem{Mark_SIC}
 M. R. Pederson, R. A. Heaton and C. C. Lin, J. Chem. Phys. {\bf 82}, 2688 (1985).



\bibitem{Jellinek}
P. H. Acioli and J. Jellinek, J. Chem. Phys. {\bf 118}, 7783 (2003).
D. P. Chong, O. V. Gritsenko, E. J. Baerends,
 J. Chem. Phys. {\bf 116} 1760  (2002).

\bibitem{Trickey}
S. B. Trickey, Phys. Rev. Lett. {\bf 56}, 881 (1986).

\bibitem{Ballone}
M. Harris and P. Ballone, Chem. Phys. Lett. {\bf 303}, 420 (1999).


\bibitem{PES}
Gengeliczki, Z; Pongor, CI; Sztaray, B
    Organometallics, {\bf 25} 2553 (2006);
Walter, M; Hakkinen, H
    E. Phys. J. D, {\bf 33} 393 (2005);
Moseler, M; Huber, B; Hakkinen, H; et al.
   Phys. Rev. B.  {\bf 68 } 165413 (2003); 
 Acioli, PH; Jellinek, J
    Phys. Rev. Lett. {\bf 89} 213402  (2002).



\bibitem{Dunlap}
 B. I. Dunlap and S. P. Karna, {\em Nonlinear Optical Materials}, Eds. S. P. Karna and A. T. Yates,
ACS Symposium Series, {\bf 628}, 164 (1996).

%\bibitem{FF1}
\end{thebibliography}
\end{document}